\documentclass[fleqn,twoside]{article}
\usepackage[headings]{espcrc2}

\readRCS
$Id: espcrc2.tex,v 1.2 2004/02/24 11:22:11 spepping Exp $
\usepackage{graphicx}
\usepackage[figuresright]{rotating}

\newcommand{\AmS}{{\protect\the\textfont2
  A\kern-.1667em\lower.5ex\hbox{M}\kern-.125emS}}

\newcommand{\bdkm}{$B^{-}\to DK^{-}$}
\newcommand{\bdkp}{$B^{+}\to DK^{+}$}

\newcommand{\bdskp}{$B^{+}\to D^{*}K^{+}$}

\newcommand{\bdksp}{$B^{+}\to DK^{*+}$}

\newcommand{\bddsks}{$B^{\pm}\to D^{(*)}K^{(*)\pm}$}

\newcommand{\bddsksp}{$B^{+}\to D^{(*)}K^{(*)+}$}

\newcommand{\dsdpim}{$D^{*-}\to \overline{D}{}^0\pi^{-}$}
\newcommand{\dsdpims}{$D^{*-}\to \overline{D}{}^0\pi^{-}$}

\newcommand{\dkpp}{$\overline{D}{}^0$$\to$$K^0_S\pi^+\pi^-$}

\title{Measurement of $\phi_3$ with Dalitz Plot Analysis of \bddsksp\  Decay}

\author{A. Garmash\address[MCSD]{Department of Physics, Princeton University, 
        Princeton, New Jersey 08544, USA}}

\begin{document}

\begin{abstract}
Results from the Belle and BaBar experiments on measurement of the weak angle
$\phi_3$ using a 
Dalitz plot analysis of the $K^0_S\pi^+\pi^-$ decay of the neutral $D$ meson 
from the \bddsks\ process are presented. The method employs the interference
between $D^0$ and $\overline{D}{}^0$ to extract the angle $\phi_3$, strong
phase $\delta$ and the ratio $r$ of suppressed and allowed amplitudes.
\end{abstract}

\maketitle

\section{Introduction}

Determinations of the Cabibbo-Kobayashi-Maskawa  matrix elements provide
important checks on the consistency of the standard model and ways to search
for new physics. The possibility of observing direct $CP$ violation in
$B\to DK$ decays was first discussed by I.~Bigi, A.~Carter and
A.~Sanda~\cite{bigi}. Since then, various methods to measure the weak angle
$\phi_3$ (also known as $\gamma$) using $B\to DK$ decays have been proposed.
All these methods are based on two key
observations: neutral $D^{0}$ and $\overline{D}{}^0$ mesons can decay to a
common final state, and the decay \bdkp\ can produce neutral $D$ mesons of
both flavors via $\bar{b}\to \bar{c}u\bar{s}$ and $\bar{b}\to \bar{u}c\bar{s}$
transitions, with a relative phase $\theta_+$ between the two interfering
amplitudes that is the sum, $\delta + \phi_3$, of strong and weak interaction
phases.  For the decay \bdkm, the relative phase is $\theta_-=\delta-\phi_3$,
so both phases can be extracted from measurements of such charge conjugate $B$
decay modes. (Unless stated otherwise charge conjugation is implied throughout
this report.) However, the use of branching fractions alone requires additional
information to obtain $\phi_3$. This is provided either by determining the
branching fractions of decays to $CP$ eigenstates (GLW method \cite{glw})
or by using different neutral $D$ final states (ADS method \cite{ads}).

A Dalitz plot analysis of a three-body final state of the $D$ meson allows
one to obtain all the information required for determination of $\phi_3$ in
a single decay mode. Three body final states such as $K^0_S\pi^+\pi^-$ have
been suggested~\cite{giri,binp_dalitz} as promising modes for the extraction
of $\phi_3$.  In the Wolfenstein parameterization of the CKM matrix elements,
the weak parts of the amplitudes that contribute to the decay \bdkp\ are
given by $V_{cb}^*V_{us\vphantom{b}}^{\vphantom{*}}\sim A\lambda^3$ (for the
$\overline{D}{}^0 K^+$ final state) and
$V_{ub}^*V_{cs\vphantom{b}}^{\vphantom{*}}\sim A\lambda^3(\rho+i\eta)$ (for
$D^0 K^+$). The two amplitudes interfere as the $D^0$ and $\overline{D}{}^0$
mesons decay into the same final state $K^0_S \pi^+ \pi^-$; the admixed state
is denoted as $\tilde{D}_\pm$. Assuming no $CP$ asymmetry in neutral $D$
decays, the amplitude of the $\tilde{D}_\pm$ decay as a function of Dalitz
plot variables $m^2_+=m^2_{K^0_S\pi^+}$ and $m^2_-=m^2_{K^0_S\pi^-}$ is 
\begin{equation}
  M_\pm=f(m^2_\pm, m^2_\mp)+re^{\pm i\phi_3+i\delta}f(m^2_\mp, m^2_\pm), 
\end{equation}
where $f(m^2_+, m^2_-)$ is the amplitude of the \dkpp\ decay, and $r$ is the
ratio of the magnitudes of the two interfering amplitudes. The value of $r$
is given by the ratio of the CKM matrix elements 
$|V_{ub}^*V_{cs\vphantom{b}}^{\vphantom{*}}|/
 |V_{cb}^*V_{us\vphantom{b}}^{\vphantom{*}}|$
and the color suppression factor, and is estimated to be in the range
0.1--0.2~\cite{gronau}.

The method has a two-fold ambiguity: the $(\phi_3,\delta)$ and
$(\phi_3+180^{\circ}, \delta+180^{\circ})$ solutions cannot be separated.
The solution with $0<\phi_3<180^{\circ}$ is chosen. 

The method described above can be applied to other $B$ decay modes such as
\bdskp\ and \bdksp.

\section{Data Samples}

Results from the two $B$-factories Belle/KEKB and BaBar/PEPII are available.
The current proceedings are based on results reported in 
Refs.~\cite{belle_phi3,babar_phi3}. For the most recent updates see
Ref.~\cite{HFAG}.
The Belle collaboration uses a data sample that consists of $386\times10^6$
$B\bar{B}$ pairs. The decay chains \bdkp, \bdskp\ with $D^*$$\to$$D\pi^0$ and
\bdksp\ with  $K^{*+}$$\to$$K^0_S\pi^+$ are selected for the analysis. 
Analysis by the BaBar collaboration is based on $227\times 10^6$ $B\bar{B}$
pairs. The reconstructed final states are \bdkp and \bdskp with two $D^*$
channels: $D^*$$\to$$D\pi^0$ and $D^*$$\to$$D\gamma$. The neutral $D$ meson is
reconstructed in the $K^0_S\pi^+\pi^-$ final state in all cases. The dominant
backgrounds come from a random combination of a real or fake $D^{(*)0}$ meson
with a charged track in continuum events; from $B^+\to D^{(*)}\pi^+$ events
with $\pi-K$ misidentification or other $B\bar{B}$ decays. 

\begin{table}
\caption{Fit results for \dkpp\ decay~\cite{belle_phi3}.}
\label{dkpp_table}
\begin{tabular}{lcc} \hline
Channel~~~~~~~~~~~~ 	     & Phase ($^{\circ}$) 
			     & Fit fraction
			     \\ \hline
$K^0_S \sigma_1$             & $212\pm 3$      & 9.8\%   \\
$K^0_S\rho^0$                & 0 (fixed)       & 21.6\%  \\
$K^0_S\omega$                & $110.8\pm 1.6$  & 0.4\%   \\
$K^0_S f_0(980)$             & $201.9\pm 1.9$  & 4.9\%   \\
$K^0_S \sigma_2$             & $237\pm 11$     & 0.6\%   \\
$K^0_S f_2(1270)$            & $348\pm 2$      & 1.5\%   \\
$K^0_S f_0(1370)$            & $82\pm 6$       & 1.1\%   \\
$K^0_S \rho^0(1450)$         & $9\pm 8$        & 0.4\%   \\
$K^*(892)^+\pi^-$            & $132.1\pm 0.5$  & 61.2\%  \\ 
$K^*(892)^-\pi^+$            & $320.3\pm 1.5$  & 0.55\%  \\
$K^*(1410)^+\pi^-$	     & $113\pm 4$      & 0.05\%  \\
$K^*(1410)^-\pi^+$	     & $254\pm 5$      & 0.14\%  \\
$K_0^*(1430)^+\pi^-$         & $353.6\pm 1.2$  & 7.4\%   \\
$K_0^*(1430)^-\pi^+$         & $88\pm 4$       & 0.43\%  \\
$K_2^*(1430)^+\pi^-$         & $318.7\pm 1.9$  & 2.2\%   \\
$K_2^*(1430)^-\pi^+$         & $265\pm 6$      & 0.09\%  \\
$K^*(1680)^+\pi^-$           & $103\pm 12$     & 0.36\%  \\
$K^*(1680)^-\pi^+$           & $118\pm 11$     & 0.11\%  \\
non-resonant                 & $164\pm 5$      & 9.7\%   \\ 
\hline
\end{tabular}
\end{table}

The numbers of reconstructed signal events in the Belle's sample are
$331\pm23$, $81\pm11$ and $54\pm8$ for the \bdkp, \bdskp and \bdksp channels,
respectively. BaBar finds $282\pm20$, $90\pm11$ and $44\pm8$ signal events in
the \bdkp, $B^+\to D^*[D\pi^0]K^+$ and $B^+\to D^*[D\gamma]K^+$ respectively.

Several high statistics control samples are reconstructed in order to
cross-check the $CP$ fit results. A sample of \dsdpim\ events produced via
the $e^+e^-\to c\bar{c}$ continuum process is selected. This decay mimics the
$B^-\to D^0 K^-$ decay with $r=0$. The $B^+\to D^{(*)}\pi^+$ mode is used as
a second control sample where $r$ is expected to be approximately 0.01.

\begin{figure}[t]
\includegraphics[width=18pc]{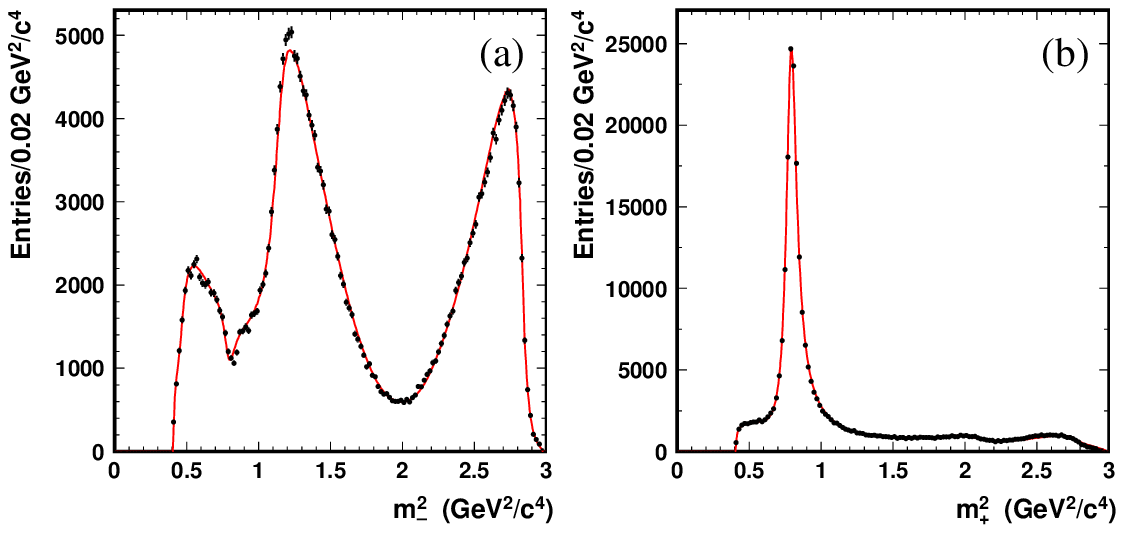}\\
\includegraphics[width=18pc]{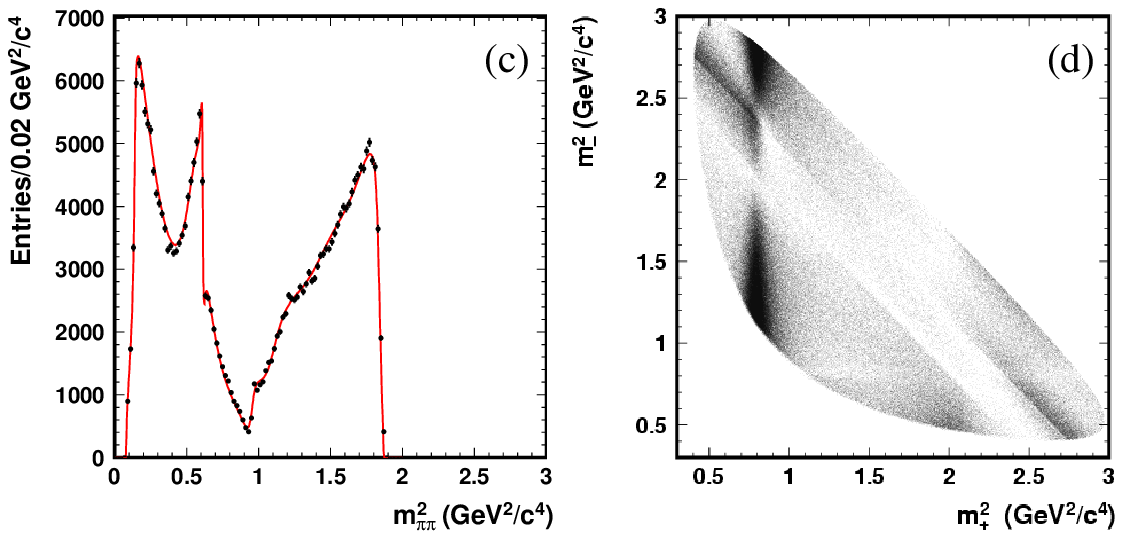}
\vspace*{-10mm}
\caption{(a) $m^2_-$, (b) $m^2_+$, (c) $m^2_{\pi\pi}$ and (d)
           Dalitz plot distribution for \dsdpims, \dkpp\ 
           decays from the $e^+e^-\to c\bar{c}$ continuum process. 
	   The points with error bars show the data;
           the smooth curve is the fit result~\cite{belle_phi3}.}
  \label{ds2dpi_plot}
\end{figure}

\section{\boldmath{\dkpp} decay amplitude}

The \dkpp\ decay amplitude $f(m^2_+, m^2_-)$ can be determined independently
from a large sample of flavor-tagged \dsdpim, \dkpp\ decays produced in
continuum $e^+e^-$ annihilation. Once $f$ is known, a fit to $B^\pm$ data
allows determination of $r$, $\phi_3$ and $\delta$. 

The amplitude $f$ is parametrized as a coherent sum of quasi-two-body
amplitudes $\mathcal{A}_j$ and a non-resonant amplitude,
\begin{equation}
  f(m^2_+, m^2_-) = \sum\limits_{j=1}^{N} a_j e^{i\alpha_j}
  \mathcal{A}_j(m^2_+, m^2_-)+
    b e^{i\beta}, 
  \label{d0_model}
\end{equation}
where $N$ is the total number of resonances and $a_j$, $\alpha_j$, $b$ and
$\beta$ are free parameters. A set of 18 quasi-two-body amplitudes is used 
to fit the data. The list of resonances included in the model as well as
parameters determined from the fit are summarized in Table~\ref{dkpp_table}.
For consistency with other analyzes \cite{babar_phi3,dkpp_cleo}, the
$\overline{D}{}^0\to K^0_S\rho^0$ mode is chosen to have unit amplitude and
zero relative phase. The \dkpp\ Dalitz plot distribution, as well as its
projections with the fit results superimposed, are shown in 
Fig.~\ref{ds2dpi_plot}.

\begin{figure}[t]
\includegraphics[width=8.7pc]{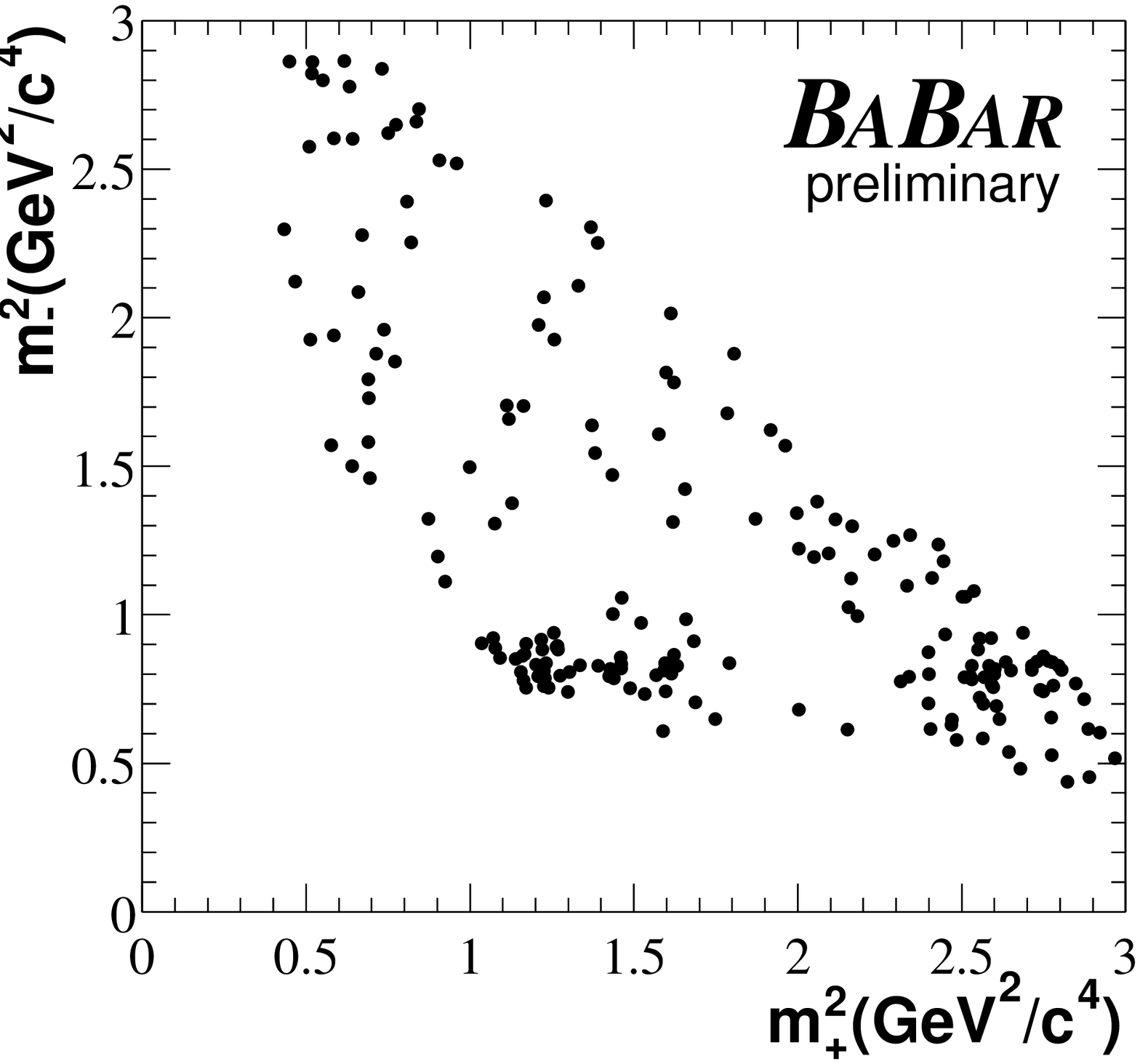}
\includegraphics[width=8.7pc]{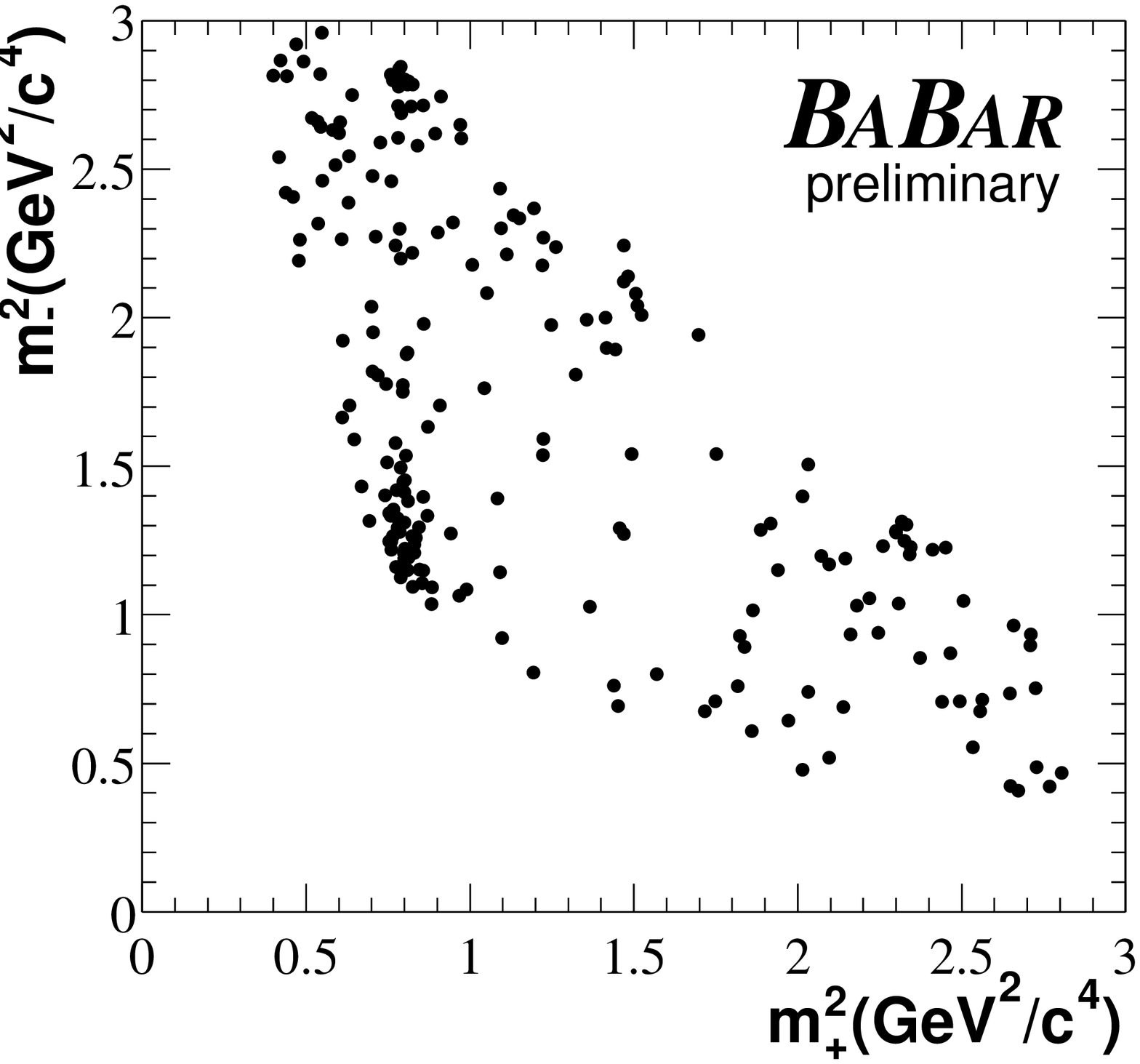}\\
\includegraphics[width=8.7pc]{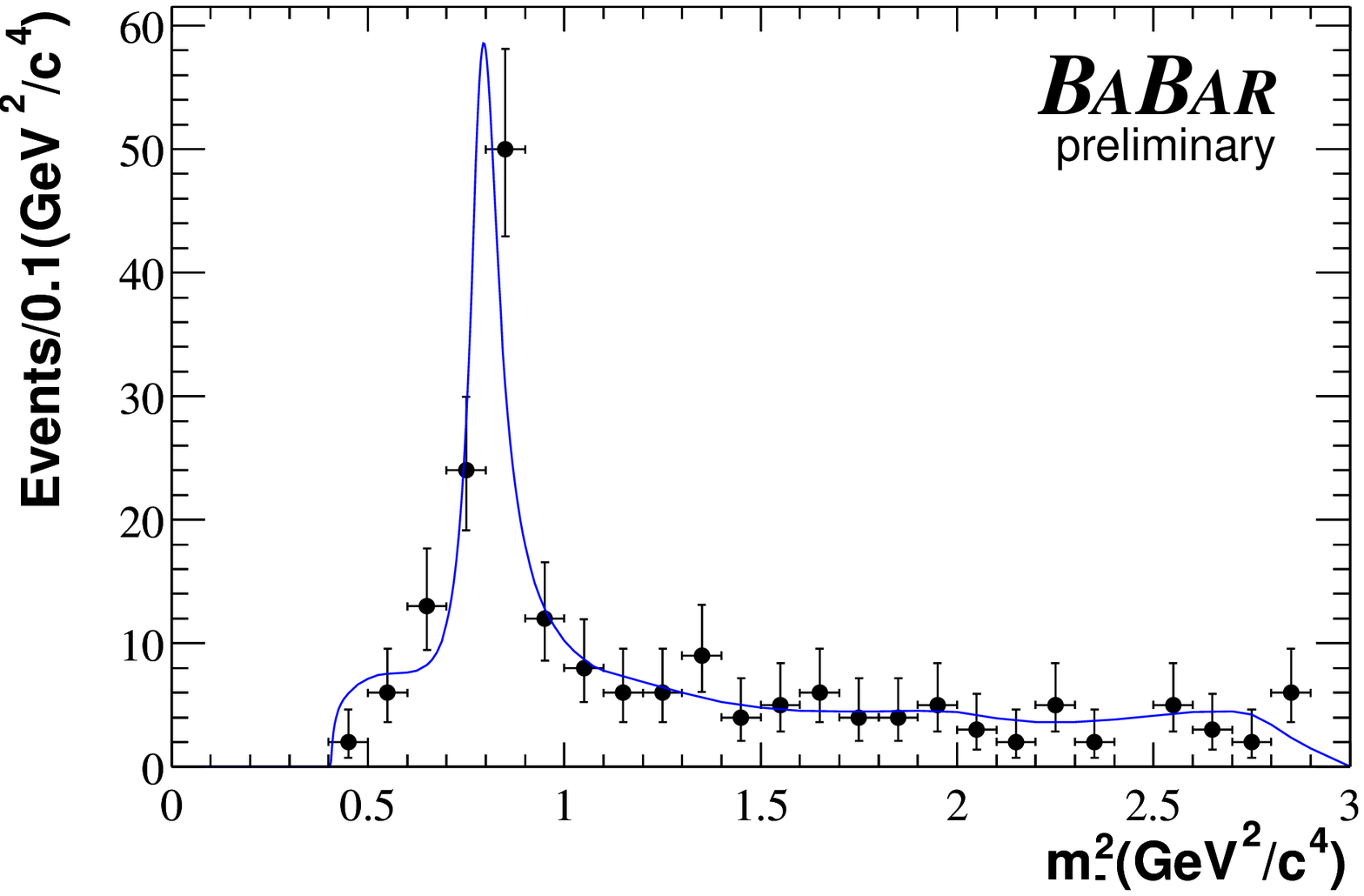}
\includegraphics[width=8.7pc]{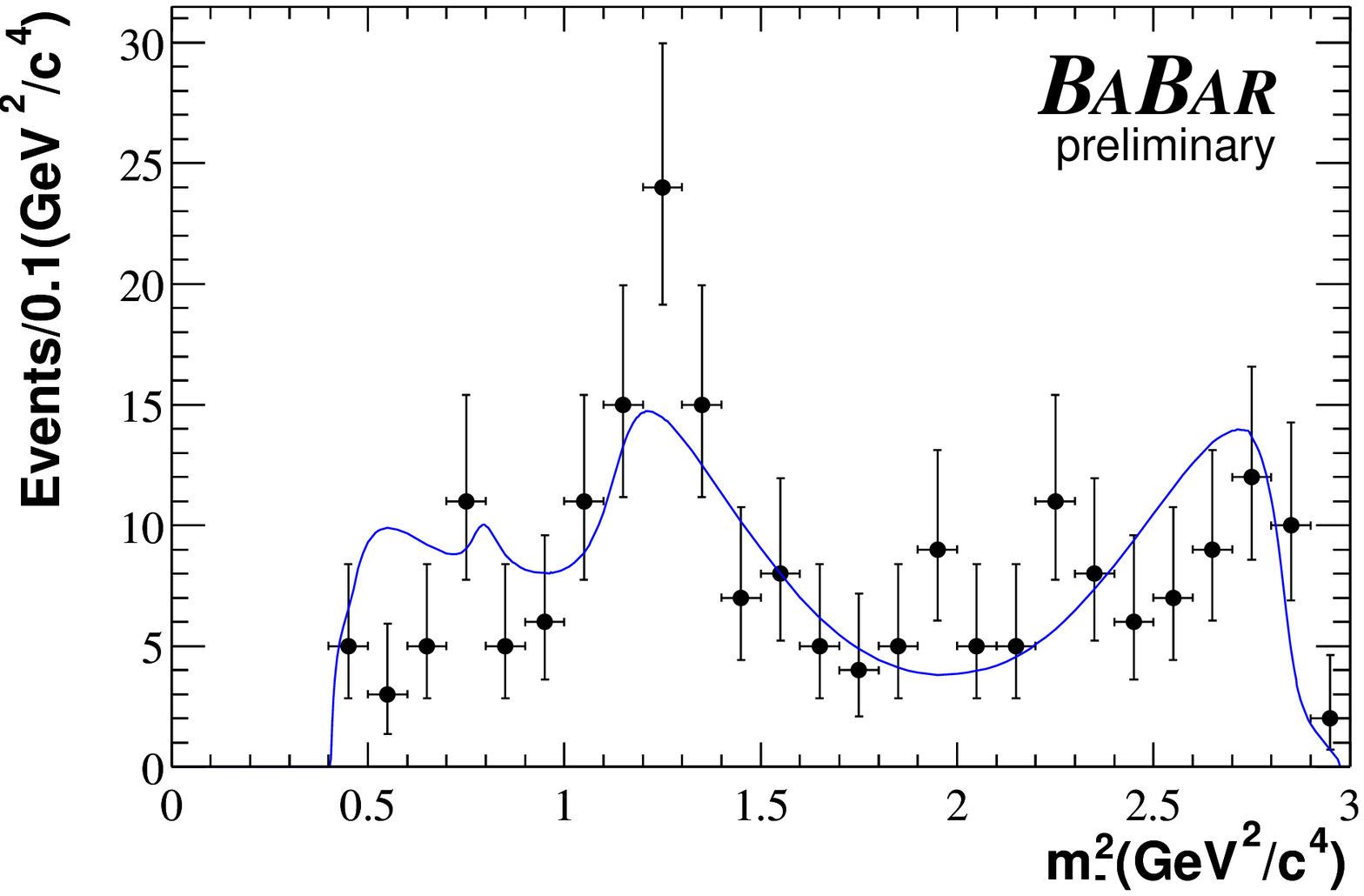}\\
\includegraphics[width=8.7pc]{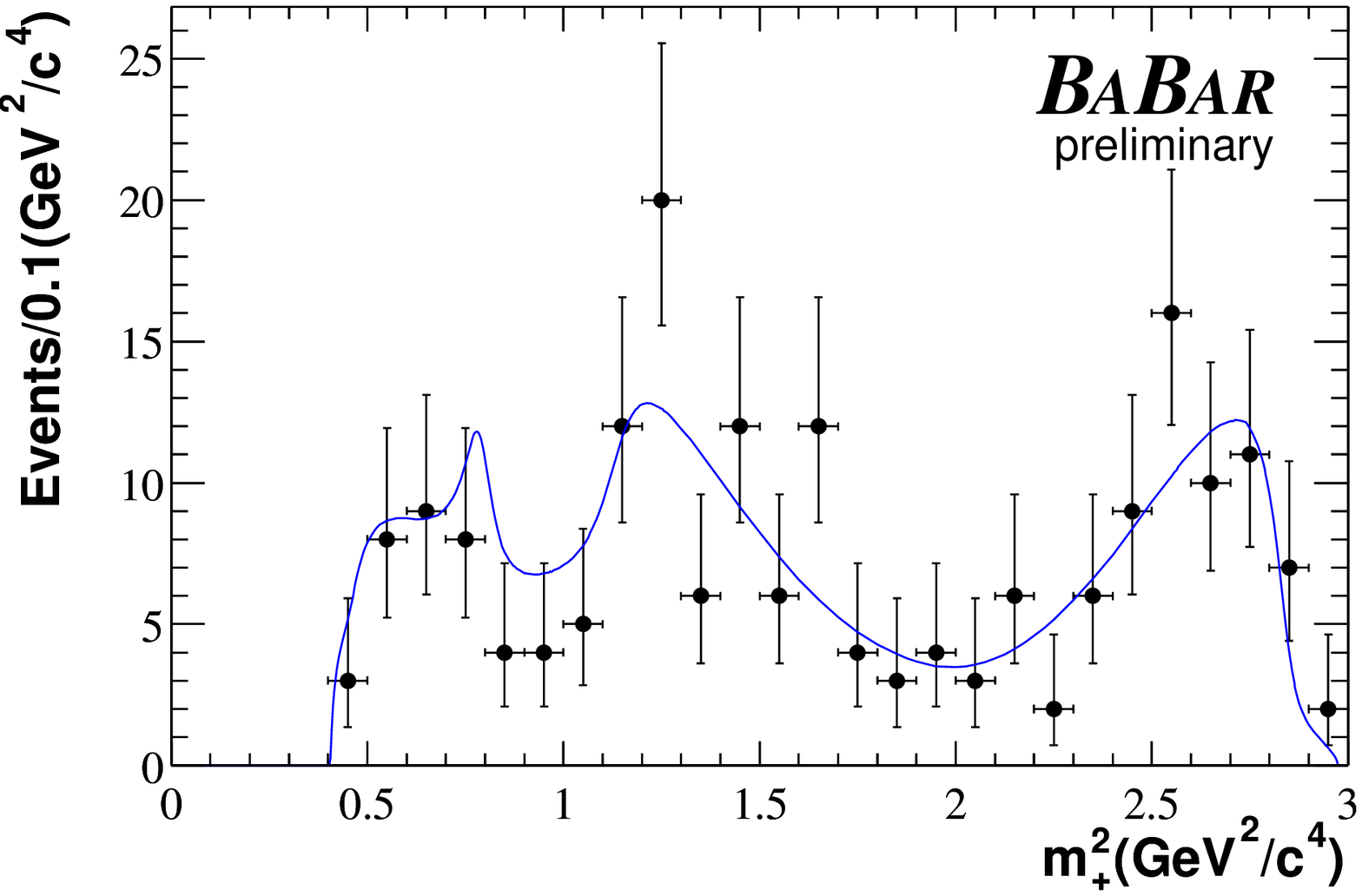}
\includegraphics[width=8.7pc]{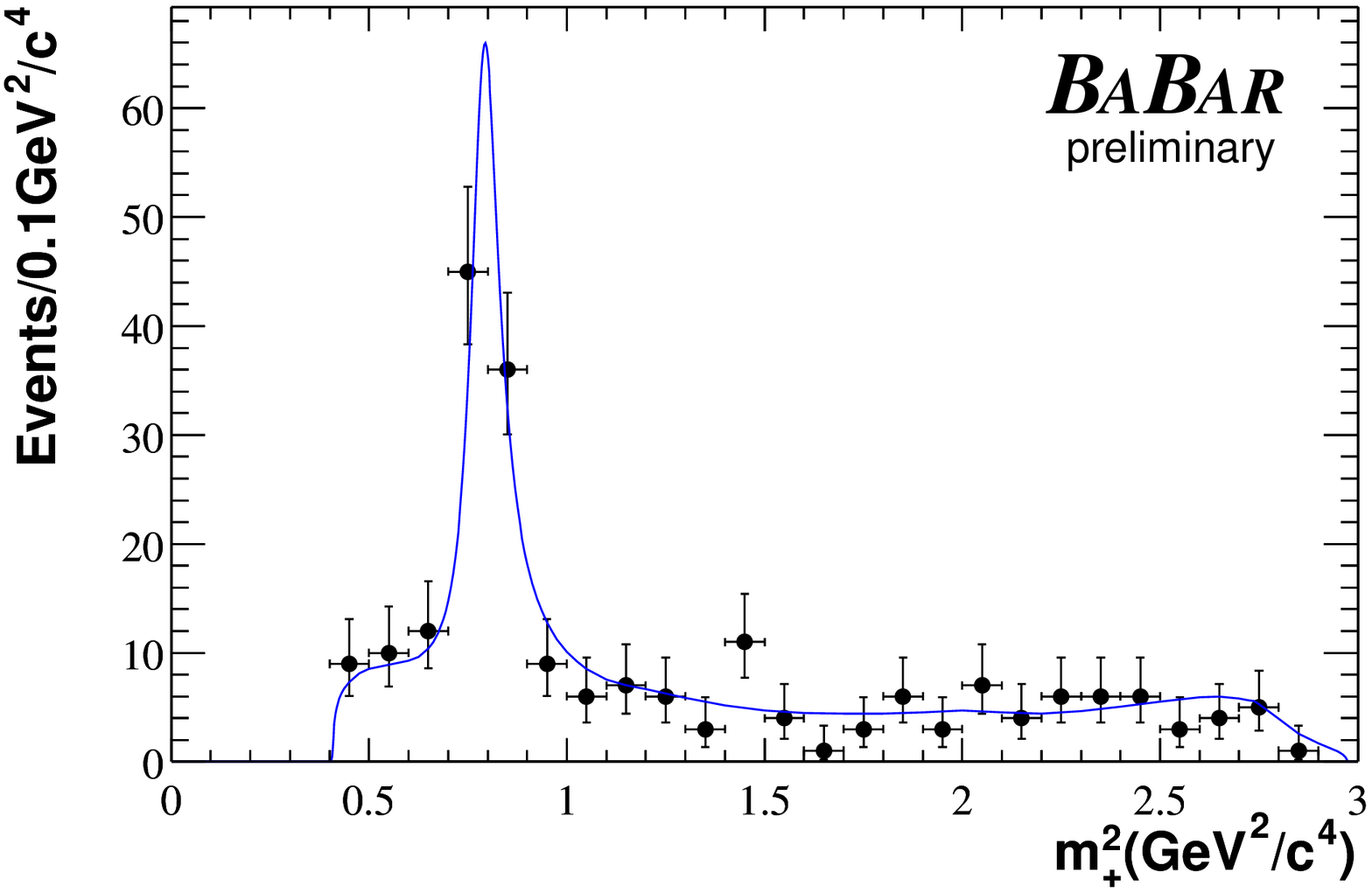}
\vspace*{-10mm}
\caption{Dalitz plot and one dimensional projections for the $B^-\to D^0 K^-$
(left column) and $B^+\to \bar{D}^0 K^+$ candidates (right column) in the $B$
signal region~\cite{babar_phi3}.}
\label{fig:DKprojdalitz}
\vspace*{-2mm}
\end{figure}
\begin{figure}[!ht]
\includegraphics[width=9.1pc]{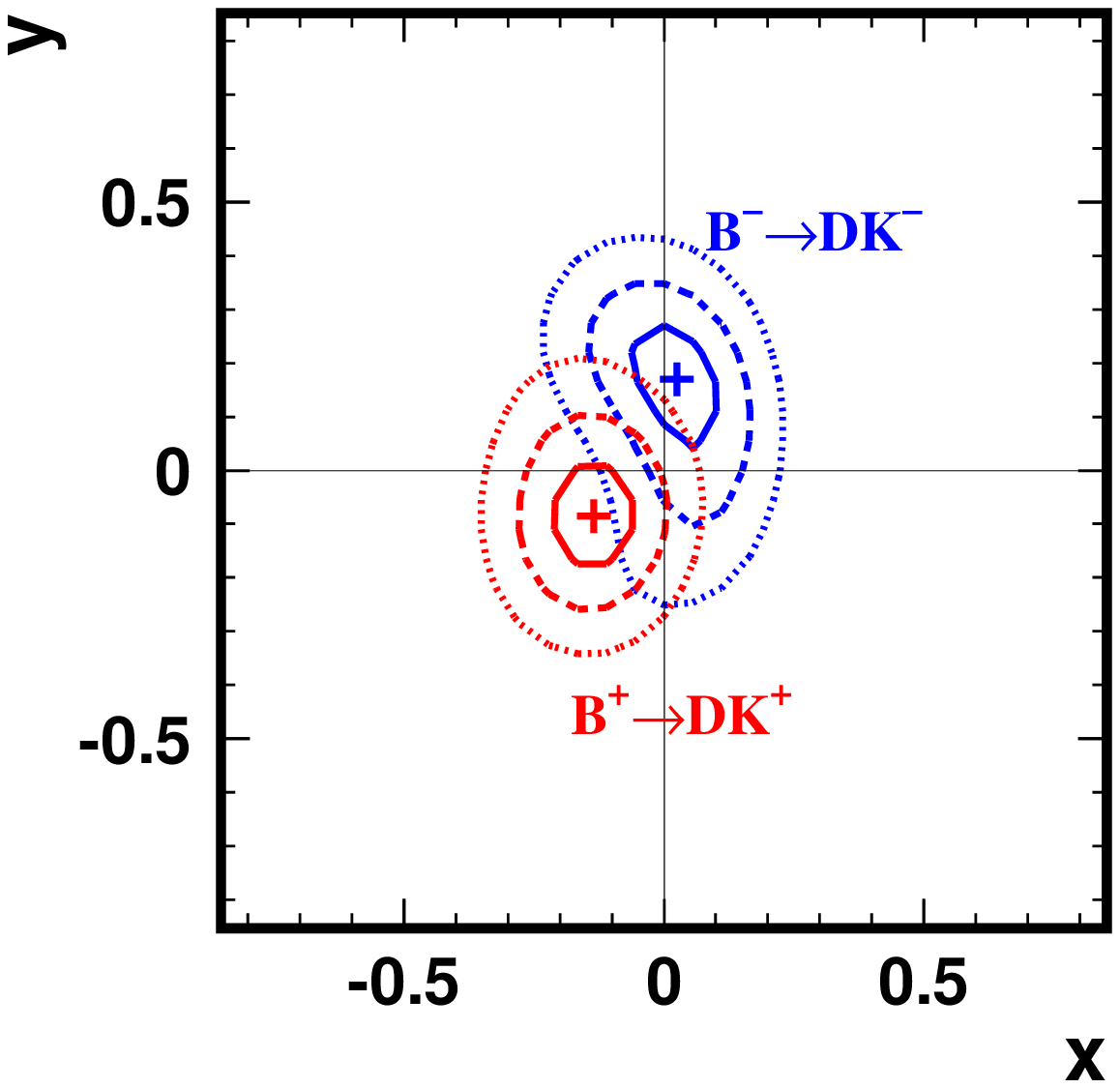} \hspace*{-4mm}
\includegraphics[width=9.1pc]{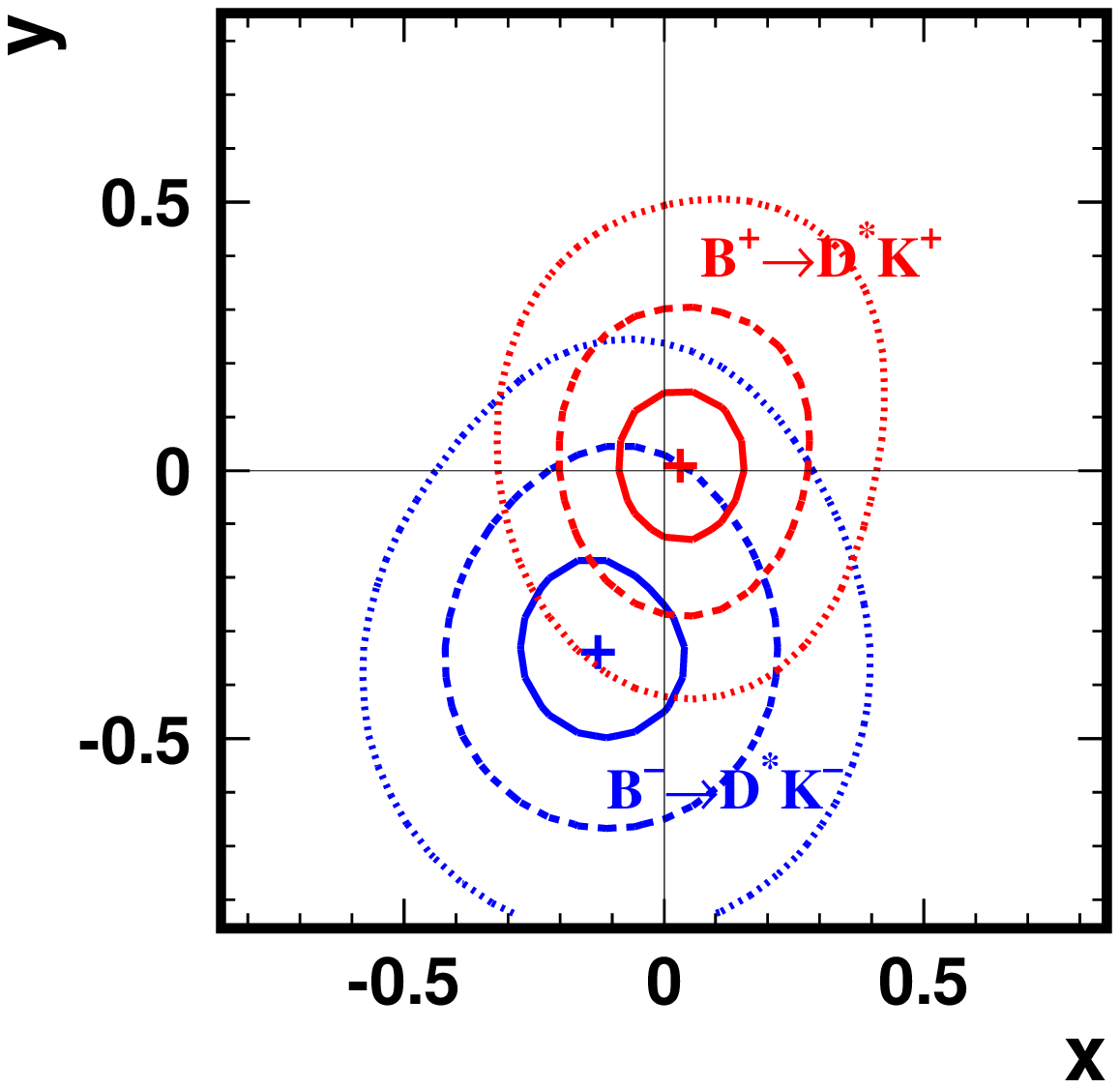}
\vspace*{-10mm}
  \caption{Results of signal fits with free parameters 
           $x=r\cos\theta$ and $y=r\sin\theta$ for (a) \bdkp and
           (b) \bdskp\ samples, separately for $B^-$
           and $B^+$ data. Contours indicate integer multiples of the 
           standard deviation~\cite{belle_phi3}. }
  \label{sig_fit}
\vspace*{-2mm}
\end{figure}

\section{$CP$ Analysis}

Analysis of $CP$ violation is performed by means of a maximum likelihood fit.
The fit is performed by minimizing the negative likelihood function
(Belle performs an unbinned fit while BaBar uses a binned approach.) 

\begin{equation}
  -2 \log L = 
  -2\sum\limits^n_{i=1}
  \log p(m^2_{+,i}, m^2_{-,i}, \Delta E_i, M_{{\rm bc},i}),
  \label{log_l_4d}
\end{equation}
with the Dalitz plot density $p$ represented as 
$$
  p(m^2_+, m^2_-, \Delta E, M_{\rm bc}) =~~~~~~~~~~~~~~~~~~~~~~~~~~~~~~~~~~
$$
$$ 
 \varepsilon|f(m^2_+, m^2_-) + (x+iy)f(m^2_-, m^2_+)|^2 F_{\rm sig}(\Delta E, M_{\rm bc})
$$
\begin{equation}
  ~~~~~~~~~~~~~~~~~~~~+F_{\rm bck}(m^2_+,m^2_-,\Delta E,M_{\rm bc}), 
  \label{dalitz_density}
\end{equation}
where $x=r\cos(\phi_3+\delta)$, $y=r\sin(\phi_3+\delta)$; $F_{\rm sig}$ is
the signal $\Delta E$ and $M_{\rm bc}$ distribution represented by the product
of two Gaussian functions (BaBar also includes Fisher discriminant in the fit);
$F_{\rm bck}$ is the distribution of the background events, and
$\varepsilon=\varepsilon(m^2_+, m^2_-)$ is reconstruction efficiency
determined from MC simulation. The background density function $F_{\rm bck}$
is determined from analysis of sideband events in data and with Monte Carlo
generated events.

The fit procedure is first tested on two high statistics control samples: 
$D^{*+}$$\to$$D^0\pi^+$ from $c\bar{c}$ continuum events and 
$B^-$$\to$$D^{(*)0}\pi^-$. In the $CP$ fit to $D^{*+}$$\to$$D^0\pi^+$ the fit
yields $r_B=(-5.2\pm5.2)\times 10^{-3}$. Results consistent with $r=0$ are
obtained with the other two control samples: $r_B=(1.8\pm1.5)\times10^{-2}$
for the $B^-$$\to$$D^{0}\pi^-$ mode and $r_{D^*K}=(4.6\pm2.1)\times10^{-2}$
for the $B^-$$\to$$D^{*0}\pi^-$ mode. This is in agreement with expectations.

\begin{table*}
  \caption{Summary of fits results. }
  \label{summary}
\begin{center}
  \begin{tabular}{lcc} \hline
  Parameter~~~~~~~~~~~ &
 ~~~~~~~~~~~~~~~~Belle~~~~~~~~~~~~~~~~ &
 ~~~~~~~~~~~~~~~~BaBar~~~~~~~~~~~~~~~~ \\ \hline

  $\phi_3$        & $53.3^{+14.8}_{-17.7}\pm2.5\pm8.7$ 
                  & $70\pm31^{+12+14}_{-10-11}$
\\
  $r_{DK}$        & $0.159^{+0.054}_{-0.050}\pm0.012\pm0.049$ 
                  & $0.12\pm0.08\pm0.03\pm0.04$
\\
  $\delta_{DK}$   & $145.7^{+19.0}_{-19.7}\pm3.0\pm22.9$
                  & $104\pm45^{+17+16}_{-21-24}$
\\
  $r_{D^*K}$      & $0.175^{+0.108}_{-0.099}\pm0.013\pm0.049$ 
                  & $0.17\pm0.10\pm0.03\pm0.03$
\\
  $\delta_{D^*K}$ & $302.0^{+33.8}_{-35.1}\pm6.1\pm22.9$
                  & $296\pm41^{+14}_{-12}\pm15$
\\
  $r_{DK^*}$      & $0.564^{+0.216}_{-0.155}\pm0.041\pm0.084$
                  & $-$
\\
  $\delta_{DK^*}$ & $242.6^{+20.2}_{-23.2}\pm2.5\pm49.3$
                  & $-$
\\ \hline
  \end{tabular}
\end{center}
\end{table*}

A summary of the results of the fits to the signal events is given in 
Table~\ref{summary}, where the first quoted error is statistical, the
second is systematic and the third is a model uncertainty. Figure~\ref{sig_fit}
demonstrates results of the fit to \bdkp\ and \bdskp\ events on the $x-y$
plane.

Systematic errors come from uncertainty in the knowledge of 
the functions used in the signal Dalitz plot fit. These include the Dalitz 
plot profiles of the backgrounds and the detection efficiency, the momentum 
resolution description, and on the parametrization of the 
$\Delta E-M_{\rm bc}$ shape of the signal and background. Though the 
statistical error is still quite large, this method currently provides
the best direct constraints on $\phi_3$.

Uncertainty in the model used to parametrize the \dkpp\ decay amplitude is 
the source of the associated error in the analysis. It arises from a non-unique
choice of the set of quasi-two-body channels as well as uncertainty in
parametrization of some components (non-resonant amplitude, for example).
To evaluate this uncertainty several alternative models have been used to fit
the data.

Although at present the $\phi_3$ accuracy is dominated by the statistical
uncertainty, the model error will eventually dominate as the experimental
statistics increase. A model independent way to extract $\phi_3$ has been
proposed in Ref.~\cite{giri}. The idea is that in addition to flavor tagged
\dkpp\ decays, one can use $CP$ tagged decays to $K^0_S\pi^+\pi^-$ from the
$\psi(3770)\to D\bar{D}$ process. Combining the two data sets the amplitude
and phase could be measured for each point on the Dalitz plot in a model
independent way. Study with Monte Carlo simulations indicates that with with
50 ab$^{-1}$ of data $\phi_3$ can be measured with a total accuracy of
better than 2 degrees.

\end{document}